\documentclass[journal,twoside]{IEEEtran}
\usepackage{cite}
\usepackage{amsmath,amssymb,amsfonts, mathtools}
\usepackage{graphicx}
\usepackage{textcomp}
\usepackage{xcolor}
\usepackage{comment}
\usepackage{commath}
\usepackage{algorithm,algorithmic}
\usepackage{enumerate}
\usepackage{psfrag,subfigure}
\usepackage{amsbsy,epsfig,bbm,mathrsfs,multirow,amsthm}
\usepackage{float}
\usepackage{mathrsfs}
\usepackage{color}
\usepackage{tabulary}
\usepackage{multirow}
\usepackage{cases}
\usepackage{stfloats}
\usepackage{url}
\usepackage[subfigure]{tocloft}
\usepackage{import}
\usepackage{xcolor}

\DeclareMathOperator{\diag}{\mathrm{diag}}
\DeclareMathOperator*{\argmax}{\mathrm{argmax}}
\DeclareMathOperator*{\clip}{\mathrm{clip}}

\begin{document}

\title{Reconfigurable Intelligent Surface-assisted Multi-UAV Networks: Efficient Resource Allocation with Deep Reinforcement Learning}

\author{Khoi Khac Nguyen, Saeed Khosravirad, Daniel Benevides da Costa, Long D. Nguyen, and Trung Q. Duong
\thanks{K. K. Nguyen and T. Q. Duong are with the School of Electronics, Electrical Engineering and Computer Science, Queen’s University Belfast, Belfast BT7 1NN, UK (e-mail: \{knguyen02,trung.q.duong\}@qub.ac.uk).}
\thanks{S. R. Khosravirad is with Nokia Bell Labs, Murray Hill, NJ 07964 USA (e-mail: saeed.khosravirad@nokia-bell-labs.com)}
\thanks{D. B. da Costa is with Future Technology Research Center, National Yunlin University of Science and Technology, Douliou, Yunlin 64002, Taiwan, R.O.C. and with the Department of Computer Engineering, Federal University of Ceara, Sobral 62010-560, CE, Brazil (e-mail: danielbcosta@ieee.org).}
\thanks{L. D. Nguyen is with Duy Tan University, Da Nang, Vietnam (e-mail: nguyendinhlong1@duytan.edu.vn).}
\thanks{This work was supported in part by the U.K. Royal Academy of Engineering (RAEng) under the RAEng Research Chair and Senior Research Fellowship scheme Grant RCSRF2021$\backslash$11$\backslash$41. }
}

\maketitle

\begin{abstract}
In this paper, we propose reconfigurable intelligent surface (RIS)-assisted unmanned aerial vehicles (UAVs) networks that can utilise both advantages of UAV's agility and RIS's reflection for enhancing the network's performance. To aim at maximising the energy efficiency (EE) of the considered networks, we jointly optimise the power allocation of the UAVs and the phase-shift matrix of the RIS. A deep reinforcement learning (DRL) approach is proposed for solving the continuous optimisation problem with time-varying channels in a centralised fashion. Moreover, parallel learning approach is also proposed for reducing the information transmission requirement of the centralised approach. Numerical results show a significant improvement of our proposed schemes compared with the conventional approaches in terms of EE, flexibility, and processing time. Our proposed DRL methods for RIS-assisted UAV networks can be used for real-time applications due to their capability of instant decision-making and handling the time-varying channel with the dynamic environmental setting.
\end{abstract}

{\it Keywords-}  Deep reinforcement learning, multi-UAV, reconfigurable intelligent surface, resource allocation.


\section{Introduction}\label{Sec:Intro}
Unmanned aerial vehicles (UAVs) are recently widely applied in numerous fields due to their agility. The high altitude of UAVs can overcome some bottlenecks of the existing scenarios, such as building blockage, remote areas, and emergency services. Some real-life applications of the UAVs are surveillance \cite{SS:19:Access}, geography exploration \cite{AC:17:Drones}, disaster rescue mission \cite{TD:19:GLOBECOM, LN:19:SPAWC,Long:EAI}, and wireless communications \cite{CZ:18:WCL, Khoi:20:Access}. The UAVs are also playing a crucial role in bringing beyond fifth generation (5G) network to every corner around the world owing to their low-cost production and flexibility. At the end, UAV-assisted wireless networks significantly enhance the network's coverage and improve the information transmit efficiency.

Very recently, reconfigurable intelligent surface (RIS) has emerged as a cutting-edge technology for beyond 5G and sixth generation (6G) networks. In particular, a massive number of reflective elements are intelligently controlled to reflect the received signal toward the destinations. The controller helps the RIS be dynamically adapted to the propagation environment with the aim to meet different purposes; for example, enhance the arrival signal and mitigate the interference \cite{HG:20:WC, YZ:20:VT, YC:21:WC, CH:19:WC, EB:19:Access, JY:21:TC, SA:20:TC,HY:20:JSAC}. The RIS has been recently deployed efficiently due to its low-cost hardware production, nearly-passive nature, easy deployment, communication without new waves, and energy-saving nature.

Owing to the intrinsic features of RIS and UAVs, the RIS-assisted UAV communications have been recently considered for enhancing network performance. Although the high altitude of the UAV significantly strengthens the channel between the UAV and the users, the connections are sometimes blocked by buildings or other obstacles in specific scenarios. Thus, the RIS attached on the building or on a high place is an option to reflect the channel from the UAV to the users \cite{LG:20:Access, SL:20:WCL, AR:21:IOT}. Moreover, the data through the RIS will experience fewer intermediate delays and more freshness than when we use a mobile active relay in the middle.  On the other hand, the RIS is easily deployed and effective in reducing power consumption.

Deep reinforcement learning (DRL) algorithms have emerged as a powerful method for an embedded optimisation and instant decision-making model in wireless networks. The DRL methods have been used for device-to-device (D2D) communication \cite{Khoi:19:Access, KK:19:Access}, UAV-assisted networks \cite{Khoi:20:Access}, and RIS-assisted wireless networks \cite{KF:20:WCL}. The neural networks are trained in the offline phase and then deployed in the terminal devices or controllers. Thus, the proper actions can be chosen in milliseconds or instant in a centralised and decentralised manner.

\subsection{Related Works}
The high-flying altitude of the UAV helps the wireless networks improve the coverage and transmit signal \cite{LN:19:SPAWC, CZ:18:WCL, Khoi:20:Access, YZ:17:WC}. In \cite{LN:19:SPAWC}, multiple UAVs were deployed in a disaster area for efficiently supporting the users. The K-means algorithm was proposed for the deployment mission, while the Block Coordinate Descent (BCD) procedure was used for maximising the worst end-to-end sum-rate. In \cite{CZ:18:WCL}, the authors used the UAV as a mobile data collector. The optimised UAV's flying path and the wake-up scheduling at the sensor nodes helped to reduce the energy consumption in both the UAV and the sensors. The authors in \cite{Khoi:20:Access} considered the UAV as an energy provider for the non-fixed power source devices to assist communications in D2D networks. In \cite{YZ:17:WC}, the UAV's trajectory was optimised to maximise the energy efficiency (EE) in an unconstrained condition and circular trajectory.

As aforementioned, RIS has been recently attracting enormous attention as an emerging technology for enabling 5G due to its unique characteristics, which include the low-cost production and less energy consumption \cite{HG:20:WC, YZ:20:VT, YC:21:WC, CH:19:WC, EB:19:Access, JY:21:TC, SA:20:TC, HY:20:JSAC, WY:20:JSAC, BD:20:JSAC}. In \cite{HG:20:WC}, an algorithm was proposed for maximising the weighted sum-rate of all users via beamforming vector and RIS phase-shift optimisation under the perfect channel state information (CSI) and imperfect CSI scenarios. In \cite{CH:19:WC}, the power allocation and the phase-shift optimisation algorithm was proposed for maximising the EE performance. In \cite{YC:21:WC}, the RIS was used for enhancing communication and reducing the interference in the D2D networks. Two sub-problems with the fixed power transmission and the discrete RIS's phase-shift matrix were considered and solved efficiently. The authors in \cite{JY:21:TC} optimised the beamforming vector at secondary users transmitter and the RIS phase-shift in a downlink multiple-input single-output (MISO) cognitive radio system with multiple RISs. The perfect CSI and imperfect CSI scenarios were considered; then, BCD procedure was used to maximise the achievable sum-rate.

By utilising both advantages of the UAV and the RIS, the network performance are significantly improved in terms of enhancing the received signal and mitigating the interference \cite{LG:20:Access, SL:20:WCL, AR:21:IOT}. In \cite{LG:20:Access}, the joint beamforming vector, trajectory and phase-shift optimisation algorithm was proposed for maximising the received signal at the ground users in the UAV-assisted wireless communications. In \cite{SL:20:WCL}, the joint UAV flying path and RIS passive beamforming design was investigated in order to maximise the network sum-rate. Two sub-problems with the fixed trajectory and the optimal phase-shift matrix were solved using a closed-form solution and the successive convex approximation method. The UAV communications supported by RIS have been extended to the concept of ultra-reliable and low-latency communication (URLLC) in \cite{AR:21:IOT} where the RIS passive beamforming, the UAV's position and URLLC blocklength were optimised for minimising the total decoding error rate of URLLC.

Sice DRL is an effective solution for solving the dynamic environment with continuous moving \cite{Khoi:19:Access, KK:19:Access, Khoi:20:Access, YY:19:SAC}, some recent works have been explored the efficiency of the DRL techniques for RIS-assisted wireless networks \cite{KF:20:WCL, BS:21:OJCS, CH:20:JSAC, MS:21:VT}. In \cite{KF:20:WCL}, a DRL algorithm was proposed for optimising the RIS phase-shift in order to maximise the signal-to-noise ratio (SNR). The author in \cite{CH:20:JSAC} optimised the transmit beamforming vector and the RIS phase-shift model by using the DRL algorithm to maximise the total sum-rate. A deep Q-learning and deep deterministic policy gradient were proposed and showed impressive results in the MISO communications. To minimise the sum age-of-information, the authors in \cite{MS:21:VT} proposed a DRL algorithm to adjust the UAV's altitude and the RIS phase-shift. However, these techniques mostly assume idealistic conditions or flat fading channel settings.

\subsection{Contributions}

However, when deploying the optimisation algorithm with DRL into RIS-assisted UAV communications, previous works assumed flat fading channels, static environment and perfect CSI, which are unrealistic and infeasible for real-life applications. Furthermore, the delay in the centralised learning and the processing time in their optimisation algorithms is huge for real-time use cases. To overcome these aforementioned shortcomings, in this paper, we propose efficient DRL algorithms by jointly optimising the power allocation of the UAV and the RIS's phase-shifts for maximising the EE and the network's sum-rate. To the best of our knowledge, our work is the first technical paper that exploits the efficiency of DRL techniques in multi-UAV-assisted wireless communications with the support of RISs. The main contributions of this work can be summarised as follows:
\begin{itemize}
	\item We conceive a wireless network of multi-UAVs supported by an RIS. Each UAV is deployed for serving a specific cluster of UEs. Due to the severe shadowing effect, the RIS is used to enhance the received signal's quality at the UEs from the associated UAV and to mitigate the interference from others.
	\item The EE problem is formulated for the downlink channel with the power restrictions and the RIS's requirement. To optimise the EE network performance, we propose a centralised DRL technique for jointly solving the power allocation at the UAVs and phase-shift matrix of the RIS. Then, parallel learning is used for training each element in our model to be intelligent.
	\item To improve the network performance, we introduce the proximal policy optimisation (PPO) algorithm with a better sampling technique.
	\item Through the numerical results, we demonstrate that our proposed methods efficiently solve the joint optimisation problem with the dynamic environmental setting and time-varying CSI and outperform the other benchmarks.
\end{itemize}

The remainder of this paper is organised as follows. We present the system model and problem formulation for the energy-efficient multi-UAV-assisted wireless communications with the support of the RIS in Section \ref{Sec:Model}. The mathematical backgrounds for the DRL algorithm are presented in Section \ref{Sec:Pre}. The centralised DDPG approach for joint optimisation of power allocation and phase-shift in multi-UAV-assisted wireless networks is introduced in Section \ref{Sec:DDPG}. We propose parallel learning for our approach to reduce delay in Section \ref{Sec:PDDPG}. Moreover, the PPO algorithm is proposed for solving both centralised and decentralised learning in Section \ref{Sec:PPO}. Numerical results are illustrated in Section \ref{Sec:Results} while the conclusion and future works are presented in Section \ref{Sec:Con}.

\section{System Model and Problem Formulation}\label{Sec:Model}
We consider a downlink multi-UAV wireless network assisted by one RIS. Each UAV is equipped with a single antenna for serving a specific cluster of a group of users (UEs), in which it is assumed $N$ UAVs corresponding to $N$ clusters of UEs, where each cluster consists of $M$ single-antenna UEs. The UEs are randomly distributed in the coverage $C$ from the centre of each cluster. The channel between the UAV and UEs is blocked by the building, wall and concretes. Thus, we deploy an RIS with $K$ elements for supporting the information transmission from UAVs to UEs.

\subsection{System Model}

We assume that the coordinate of the $n$th UAV and $m$th UEs in the $n$th cluster at the time step $t$ is $X^t_n = \Big (x^t_n, y^t_n, H^t_n \Big)$ and $X^{t}_{mn} = (x^{t}_{mn}, y^{t}_{mn})$, with $n = 1, \dots , N$ and $m =1, \dots, M$. The RIS is attached at the building or a high location at $(x^t, y^t, z^t)$, respectively.

\begin{figure}[h!]
	\centering
	\subfigure{\includegraphics[width=0.5\textwidth]{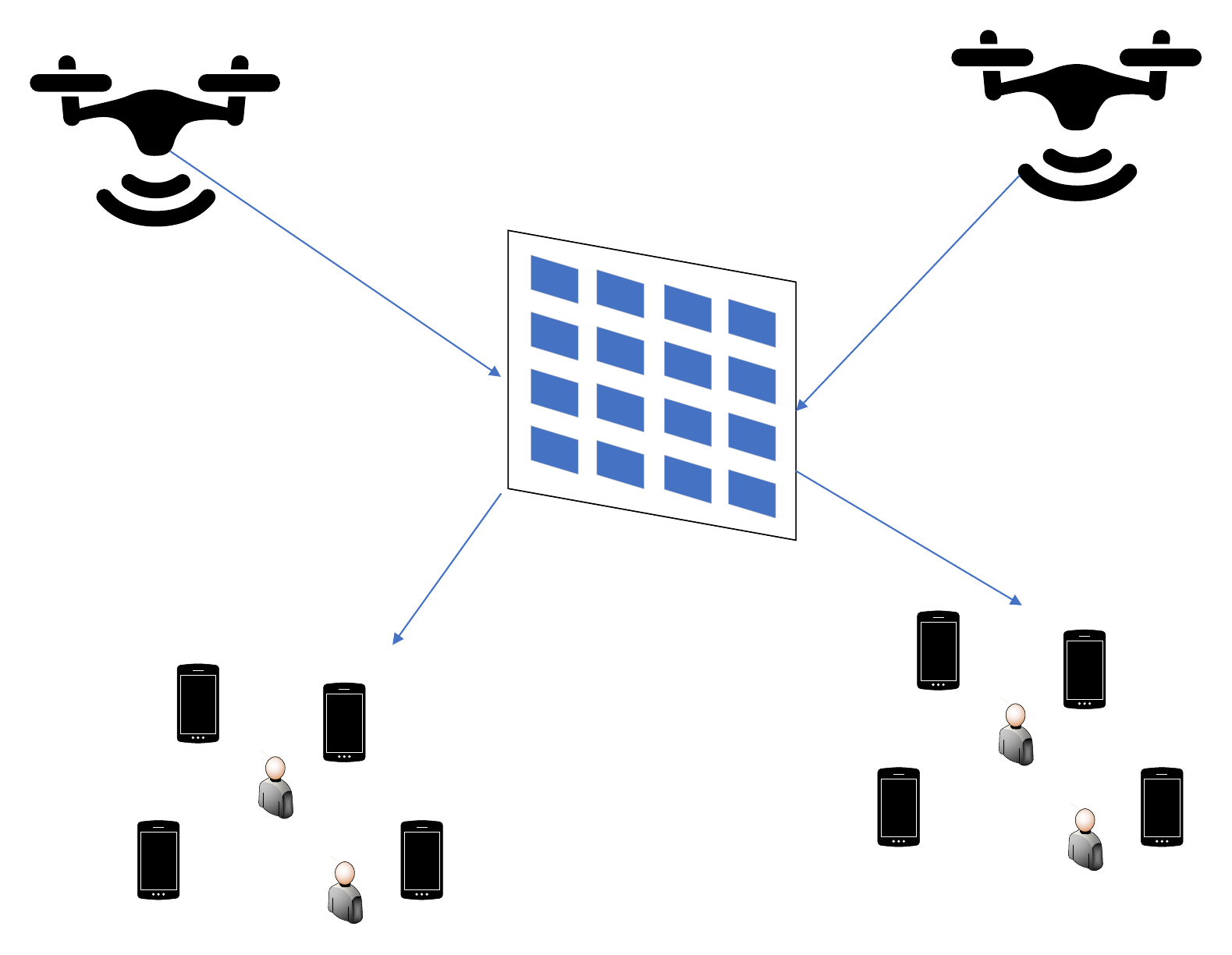}}
	\caption{System setup.}
	\label{fig:System}
\end{figure}
The distance between the $n$th UAV and the RIS panel in time step $t$ is denoted by
\begin{equation}
\begin{split}
d_{n}^t =  \sqrt {(x_n^t -x^t)^2 +  (y^t_n - y^t)^2 + (H^t_n - z^t)^2}.
\end{split}
\end{equation}

Similarly, the distance between the RIS panel and the $m$th UEs in the $n$th cluster is written as
\begin{equation}
d_{nm}^t = \sqrt {(x^t -x_{nm}^t)^2 + (y^t - y^t_{nm})^2 + (z^t)^2}.
\end{equation}

Due to the high shadowing and severe blocking effect, the direct links between UAVs and UEs do not exist and therefore it is only considered the alternative paths (reflected links) via RIS's reflection. The links between the UAVs and the RIS are modelled as air-to-air (AA) channels whereas the link between the RIS and the UEs is assumed to follow air-to-ground (AG) channel. Following the AA channel model, the channel gain between the $n$th UAV and the RIS in time step $t$ is formulated as
\begin{equation}\label{equ:H}
\begin{split}
H_{n,RIS}^t = \sqrt{\beta_0(d_{n}^t)^{- \kappa_1}} \Big[ 1, & e^{-j \frac{2\pi}{\lambda}d \cos(\phi^t_{AoA})},\\
& \dots, e^{-j \frac{2\pi}{\lambda}(K -1) d \cos(\phi^t_{AoA})} \Big]^T,
\end{split}
\end{equation}
where $\kappa_1$ is the path loss exponent for the UAV-RIS link, $d$ is element spacing, and $\lambda$ is the carrier wavelength; the right term of (\ref{equ:H}) is the signal from the $n$th UAV to the RIS, $\cos(\phi^t_{AoA})$ is the cosine of the angle-of-arrival (AoA).

According to the AG channel model, the channel gain between the RIS and the $m$th UEs in the $n$th cluster can be written as
\begin{equation}
\begin{split}
&h_{RIS, nm}^t  = \\ &\sqrt{\beta_0 (d_{nm}^t)^{-\kappa_2} } \Bigg( \sqrt{ \frac{\beta_1} {1+\beta_1}} h^{LoS}_{RIS, nm} + \sqrt{\frac{1}{\beta+1}} h^{NLoS}_{RIS, nm} \Bigg),
\end{split}
\end{equation}
where the deterministic LoS component is denoted by $ h^{LoS}_{RIS,nm} =
\big[ 1, e^{-j \frac{2\pi}{\lambda}d\cos(\phi^t_{AoD})}, \dots, e^{-j\frac{2\pi}{\lambda}(K-1)d \cos(\phi^t_{AoD})} \big]$ and the non-light-of-sight (NLoS) component is modelled as complex Gaussian distribution with a zero-mean and unit-variance $\mathcal{C} \mathcal{N} (0, 1)$;  $\cos(\phi_{AoD})$ is the angle of departure (AoD) from the RIS to the $m$th UE in the $n$th cluster; $\beta_1$ and $\kappa_2$ are the Rician factor and the path loss exponent for the RIS-UEs link, respectively.

The signal from the UAV to UEs is reflected by the RIS. Thus, the received signal from the $n$th UAV to the $m$th UE in the $n$th cluster at time step $t$ can be written as
\begin{equation}
y_{nm}^t =  H_{n, RIS}^t \Phi^t h_{RIS, nm}^t  \sqrt{P_n}x + \eta,
\end{equation}
where $H^t_n \in \mathbb{C}^{1\times K}$ is the channel gains array from the $n$th UAV to the RIS, $\eta$ is the power noise signal following the complex Gaussian distribution with power $\alpha^2$; $P_n$ and $x$ are the transmit power and the symbol signal sent from the $n$th UAV, respectively; $\Phi^t = \diag[\phi_1^t, \phi_2^t, \dots, \phi^t_K]$ is the diagonal matrix at the RIS, where $\phi_k^t = e^{j\theta^t_k}, \forall k = 1, 2, \dots, K$ with $\theta_k^t \in [0, 2\pi]$ is the phase-shift of the $k$th element in the RIS at time step $t$.

\subsection{Problem Formulation}
In this work, we consider a downlink communications where signal from the UAV is dedicated to a designated UE in the associated cluster. In other words, the $m$th UE in the $n$th cluster receives the information from the $n$th UAV while the signals from other UAVs are considered as interference. Thus, the received signal-to-interference-plus-noise-ratio (SINR) at the $m$th UE in the cluster $n$ at time step $t$ can be formulated as follows:
\begin{equation}
\gamma_{nm}^t = \frac{P_n^t | H^t_{n, RIS} \Phi^t h_{RIS, nm}^t| ^ 2}{\sum_{i \ne n}^N P_i^t | H_{i, RIS}^t \Phi^t h_{RIS, im}^t| ^2+ \alpha^2},
\end{equation}

The throughput at the $m$th UEs in the $n$th cluster at time step $t$ is written as
\begin{equation}
R_{nm}^t = B \log_2 (1 + \gamma_{nm}^t),
\end{equation}
where $B$ is the bandwidth. The total throughput at time step $t$ is cumulative from the UEs of all clusters and it can be expressed by
\begin{equation}
R_{total}^t = \sum^N_{n=1} \sum^M_{m=1} R_{nm}^t,
\end{equation}
and the total power consumption is given by
\begin{equation}
P_{total} = \sum_{n=1}^N  P_n +  P_K + P_c,
\end{equation}
where $P_K$ and $P_c$ are the power consumption at the RIS and the power circuit at the UAV, respectively.

Our objective is to maximise the EE of all UEs by jointly optimising the transmit powers at the UAVs and the phase-shifts at the RIS. In each time step $t$, each UAV will choose the proper power and each RIS's element will choose the phase-shift value depending on the local information that each component receives from the environment. The optimisation of maximising the EE of all UEs subject to the transmit power at UAVs and phase-shifts of RIS can be formulated as
 \begin{equation}\label{equ:EE}
\begin{split}
\max_{P, \Phi}  \quad& \frac{ \sum^N_{n=1} \sum^M_{m=1} R_{nm}^t}{ \sum_{n=1}^N P_n +  P_K + P_c}\\
s.t. \quad &0 \le P_n \le P_{max}, \forall n \in N, \\
& \theta_k \in [0, 2 \pi], \forall k \in K,
\end{split}
\end{equation}
where $P = \{ P_1, \dots, P_N \}$ and $P_{max}$ are the vector of power and the maximum information transmission power at the UAVs, respectively. To solve the maximised EE problem, we propose two DRL algorithms for centralised approach and then the parallel learning distributed approach is introduced for practical applications.

\section{Preliminaries}\label{Sec:Pre}
To deploy a system with the support of the DRL algorithms, we have two main approaches: value search and policy search. In the value search approach, we consider the gap between the received reward in two samples to adjust the value function. In the policy search algorithm, we directly find the policy for the problems. We represent the Markov Decision Process (MDP) \cite{BD:95:Book:v1} by $< \mathcal{S}, \mathcal{A}, \mathcal{P}, \mathcal{R}, \zeta >$, where $\mathcal{S}, \mathcal{A}$ denote the agent's state space and action space; $\mathcal{P}_{ss'}(a)$ denotes the state transition probability with $s = s^t, s' =s^{t+1} \in \mathcal{S}$, $a \in \mathcal{A}$; $r \in \mathcal{R}$ is the reward function; and $\zeta$ is the discount factor.

\subsection{Value Function}
The idea of the value function methods relies on the estimation of the value in a given state. The state-value function $V^\pi(s)$ is obtained following the policy $\pi$ starting at the state $s$ as
\begin{equation}
V^\pi = \mathbb{E} \Big \{\mathcal{R}|s, \pi \Big\},
\end{equation}
where $\mathbb{E} \{ \cdot \}$ is the expectation operation that depends on the transition function $\mathcal{P}_{ss'(a)}=p(s'|s,a)$  and the stochastic property of the policy $\pi$.

Our goal is to find the optimal policy $\pi^*$, which has a corresponding to the optimal state-value function $V^*(s)$ as
\begin{equation}
V^*(s) = \max_\pi V^\pi(s), s \in \mathcal{S}.
\end{equation}

To maximise the expected cumulative reward, the agent chooses the action $a \in \mathcal{A}$ following the optimal policy $\pi^*$ that satisfies the  Bellman equation \cite{BD:95:Book:v1}
\begin{equation}
V^* (s) = V ^{\pi^*} = \max_{a \in \mathcal{A}} \Bigg \{ \mathbb{E}\Big(r(s,a)\Big) + \zeta \sum_{s' \in \mathcal{S}}{P_{ss'} (a)V^*(s')} \Bigg \} .
\end{equation}

The action-value function is defined as the obtained reward when the agent takes action $a$ at the state $s$ under the policy $\pi$ as
\begin{equation}
Q ^\pi (s, a) = \mathbb{E} \Big(r(s, a)\Big) + \zeta \sum_{s' \in \mathcal{S}}{P_{ss'}(a) V(s')}.
\end{equation}

The optimal policy $Q ^*(s,a) = Q^{\pi^*}$, we have
\begin{equation}
V ^*(s) = \max_{a \in \mathcal{A}} Q^*(s,a)
\end{equation}

\subsection{Policy Search}
Instead of considering the value function model, the agent can directly find an optimal policy $\pi^*$. Among policy search methods, the policy gradient is most popular due to its efficient sampling with a large number of parameters. The reward function is defined by the performance under the policy $\pi$ as
\begin{equation}
J(\theta) = \sum_{s \in \mathcal{S}} d^\pi (s) \sum_{a \in \mathcal{A}} \pi_\theta(a|s) r^\pi (s,a),
\end{equation}
where $\theta_\pi$ is the vector of the policy parameters and $d^\pi(s)$ is the stationary distribution of Markov chain with the policy $\pi_\theta$. The optimal policy $\pi^*$ can be obtained by using gradient ascent for adjusting the parameters $\theta_\pi$ relying on the $\nabla_\theta J(\theta_\pi)$. For any MDP, we have \cite{sutton2000policy}
\begin{equation}
\begin{split}
\nabla_\theta J &= \sum_{s \in \mathcal{S}} d^\pi (s) \sum_{a \in \mathcal{A}} \nabla_\theta \pi(a|s) Q^\pi (s,a)\\
& =\mathbb{E}_{\pi_\theta}\Big[\nabla_\theta \ln \pi_\theta(s,a) Q^\pi(s,a) \Big]
\end{split}
\end{equation}

The REINFORCE algorithm, a Monte-Carlo policy gradient learning, adjusts the parameters $\theta_\pi$ by estimating the return using Monte-Carlo methods and episode samples. The optimal policy parameter $\theta^*_\pi$ can be obtained by
\begin{equation}
\theta^*_\pi = \argmax_{\theta_\pi} \mathbb{E} \left[\sum_{a} \pi(a|s; \theta_\pi) r(s,a) \right],
\end{equation}

The gradient is defined as
\begin{equation}
\nabla \theta_\pi = \mathbb{E}_\pi \Big [\nabla_{\theta_\pi} \ln\pi(a|s; \theta_\pi)r (s,a)|_{s=s^t, a= a^t} \Big].
\end{equation}

We use the gradient ascent to update the parameters $\theta_\pi$ as
\begin{equation}
\theta_\pi \leftarrow \theta_\pi + \varepsilon \nabla \theta_\pi,
\end{equation}
where $0 \le \varepsilon \le 1$ is the step-size parameter. The optimal action $a^*$ can be obtained with the maximum probability as follows:
\begin{equation}
a^* = \argmax_a \pi(a|s; \theta_\pi).
\end{equation}


\section{Centralised Optimisation for Power Allocation and Phase-shift Matrix}\label{Sec:DDPG}
In the centralised approach, we assume that the information is processed at a central point (e.g., cloud server) and the next action for each element in the system will be transferred at the beginning of each time step. Thus, for jointly optimising the power allocation at the UAVs and the phase-shift matrix at the RIS, we consider the central processing point as an agent. The optimisation problem can be formulated by the MDP $< \mathcal{S}, \mathcal{A}, \mathcal{P}, \mathcal{R}, \zeta >$. Particularly, with our centralised optimisation, we formulate the game as follows:

\begin{itemize}
	\item{\emph{State space}}: The agent interacts with the environment for maximising the EE performance. Thus, the agent only has knowledge about the local information, e.g., the reflected channel gains. The state space is defined as follows:
	\begin{equation}\label{equ:state}
	\begin{split}
	\mathcal{S} = & \{ H_{1,RIS} \; \Phi \; h_{RIS, 11}, \:  H_{1,RIS} \; \Phi \; h_{RIS, 12}, \\ &\dots, \: H_{n, RIS} \;\Phi \;h_{RIS, nm}, \:\dots,\: H_{N, RIS} \;\Phi\; h_{RIS, NM} \}.
	\end{split}
	\end{equation}
	
	\item{\emph{Action space}}: With the downlink transmission in the RIS-assisted multi-UAV networks, we optimise the power allocation at UAVs and phase-shift matrix at RIS. Thus, the action space is defined as follows:
	\begin{equation}
	\mathcal{A} = \{ P_1, P_2, \dots, P_N, \theta_1, \theta_2, \dots, \theta_K \}.
	\end{equation}
	
	The agent takes the action $a^t = \{ P_1^t, P^t_2, \dots, P_N^t, \theta_1^t, \theta_2^t, \dots, \theta_K^t \}$ at the state $s^t$ and moves to the next state $s' = s^{t+1}$.

	\item{\emph{Reward function}}: Our objective is to maximise the EE performance; thus, we formulate the reward function as
	\begin{equation}\label{equ:reward}
	\mathcal{R} = \frac{ \sum^N_{n=1} \sum^M_{m=1} R_{nm}^t}{ \sum_{n=1}^N P_n +  P_K+P_c}.
	\end{equation}
\end{itemize}
After formulating the EE game, we proposed a DRL algorithm for the agent to interact with the environment to find the optimal policy $\pi^*$. Deep deterministic policy gradient (DDPG) is a hybrid model composed of the actor part based on value function and the critic component based on the policy search. In the DDPG algorithm, we use \emph{experience replay buffer} and \emph{target network} techniques to improve the convergence speed and avoid excessive calculation. In the \emph{experience replay buffer}, we use a finite size of a memory size $B$ to store the executed transition $<s^t, a^t, r^t, s^{t+1}>$. After collecting enough samples, we randomly select a mini-batch $D$ of transitions from buffer $B$ for training the neural networks. The memory $B$ is set to a finite size for updating the new sample and discarding the old ones. Otherwise, we use \emph{target networks} for the critic and actor network when calculating the target value.

We denote the critic network as $Q(s, a; \theta_q )$ with the parameter $\theta_q$ and the target critic network as $Q'(s,a ;\theta_{q'})$ with the parameter $\theta_{q'}$. Similarly, we initialise the actor network $\mu(s; \theta_{\mu})$ with the parameter $\theta_{\mu}$ and the target actor network $\mu'(s; \theta_{\mu'})$ with the parameter $\theta_{\mu'}$. We train the actor and critic network using the stochastic gradient descent (SGD) over a mini-batch of $D$ samples. The critic network is updated by minimising
\begin{equation}
\label{equ:loss}
L = \frac{1}{D}\sum_{i}^D \Bigg(y^i - Q(s^i, a^i; \theta_{q}) \Bigg)^2,
\end{equation}
with the target
\begin{equation}
\begin{split}
y^i = r^i(s^i,a^i) + \zeta Q'(s^{i+1}, a^{i+1}; \theta_{{q'}})|_{a^{i+1} = \mu'(s^{i+1}; \theta_{\mu'})}.
\end{split}
\end{equation}

The actor network parameters are updated by
\begin{equation}
\label{equ:updateActor}
\nabla_{\theta_{\mu}}J \approx \frac{1}{D}\sum_{i}^D \nabla_{a^i} Q(s^i, a^i; \theta_{q})|_{a^i=\mu(s^i)} \nabla_{\theta_{\mu}} \mu(s^i; \theta_{\mu}).
\end{equation}

The target actor network parameters $\theta_{q}$ and the target critic network parameters $\theta_{\mu'}$ are updated by using soft target updates as follows:
\begin{equation}
\label{equ:updatePar1}
\theta_{q'} \leftarrow \varkappa\theta_{q} + (1-\varkappa) \theta_{q'},
\end{equation}
\begin{equation}
\label{equ:updatePar2}
\theta_{\mu'} \leftarrow \varkappa \theta_{\mu} + (1-\varkappa) \theta_{\mu'}.
\end{equation}
where $\varkappa$ is a hyperparameter between $0$ and $1$.

In the DDPG algorithm, the deterministic policy is trained in an off-policy way; thus, for \emph{explorations} and \emph{explotations} purpose, we add a noise process of $\mathcal{N}(0,1)$ as follows \cite{Lillicrap:15}:
\begin{equation}
\mu'(s^t; \theta^t_{\mu'}) = \mu(s^t; \theta^t_{\mu}) + \psi \mathcal{N}(0,1)
\end{equation}
where $\psi$ is a hyperparameter. The details of our DDPG algorithm-based technique for joint power allocation and phase-shift matrix optimisation in RIS-assisted UAV communications are presented in Algorithm \ref{alg:DDPG}, where $E$ and $T$ denote the number of the maximum episode and time step, respectively.
 \begin{algorithm}[t!]
 	\caption{Centralised optimisation for joint power allocation and phase-shift matrix in RIS-assisted UAV communications.}
 	\begin{algorithmic}[1]
 		\label{alg:DDPG}
 		\STATE Initialise the critic network $Q(s, a; \theta_q)$ and the target critic networks $Q'$
 		\STATE Initialise the actor network $\mu(s; \theta_\mu)$ and the target actor network $\mu'$
 		\STATE Initialise replay memory pool $\mathcal{B}$
 		\FOR{episode = $1,\dots, E$}
 		\STATE Initialise an action exploration process $\mathcal{N}$
 		\STATE Receive initial observation state $s^0$
 		\FOR{iteration = $1,\dots, T$}
 		\STATE Execute the action $a^t$ obtained at state $s^t$
 		\STATE Update the reward $r^t$ according to (\ref{equ:reward})
 		\STATE Observe the new state $s^{t+1}$
 		\STATE Store transition $(s^t, a^t, r^t, s^{t+1})$ into replay buffer $\mathcal{B}$
 		\STATE Sample randomly a mini-batch of $D$ transitions $(s^i, a^i, r^i, s^{i+1})$ from $\mathcal{B}$
 		\STATE Update critic parameter by stochastic gradient descent using loss function in (\ref{equ:loss})
 		\STATE Update the actor policy parameter in (\ref{equ:updateActor})
 		\STATE Update the target networks as in (\ref{equ:updatePar1}) and (\ref{equ:updatePar2})
 		\STATE Update the state $s^t = s^{t+1}$
 		\ENDFOR
 		\ENDFOR
 	\end{algorithmic}
 \end{algorithm}

\section{Parallel DRL for Joint Power Allocation and Phase-shift Matrix Optimisation}\label{Sec:PDDPG}
In practical applications, when we process all the data in a centralised manner, the information of the UAV's power and the RIS's phase-shift for the next action need to transfer at the beginning of each time step. The delay will be occurred and make the system unable to deal efficiently with the dynamic environment. Thus, we propose a parallel DRL (PDRL) technique for joint power allocation and phase-shift matrix optimisation. As the definition of the DRL model, the agents do not know the environmental factor. Thus, in our system, the $n$th UAV has no idea about the power of the $m$th UAV and the diagonal matrix at the RIS. Similarly, the RIS controller does not know about the transmit power at the UAV.

To make the UAV and the RIS work cooperatively, we consider a multi-agent learning for our system. In particular, each UAV acts as an agent and the RIS is a separated agent. For all the agents, we define the state space as $\mathcal{S} = \{ H_{1,RIS} \; \Phi\; h_{RIS, 11}, \: H_{1,RIS} \;\Phi \; h_{RIS, 12}, \: \dots, \newline \: H_{n, RIS} \; \Phi \; h_{RIS, nm}, \: \dots, \: H_{N, RIS} \; \Phi \; h_{RIS, NM} \}$ with respect to the channel state information, i.e., the compound of channel gains and phase-shifts of RIS. The UAV and the RIS process independelty, thus, the action space for the $n$th UAV agent is the transmit power $\mathcal{A}_n =\{ P_n\}$ and for the RIS agent is the phase-shift matrix $\mathcal{A}_{RIS} =\{ \theta_1, \theta_2, \dots, \theta_K \}$. With the rewards function, we rely on \eqref{equ:reward}.

\begin{figure}[h!]
	\centering
	\subfigure{\includegraphics[width=0.5\textwidth]{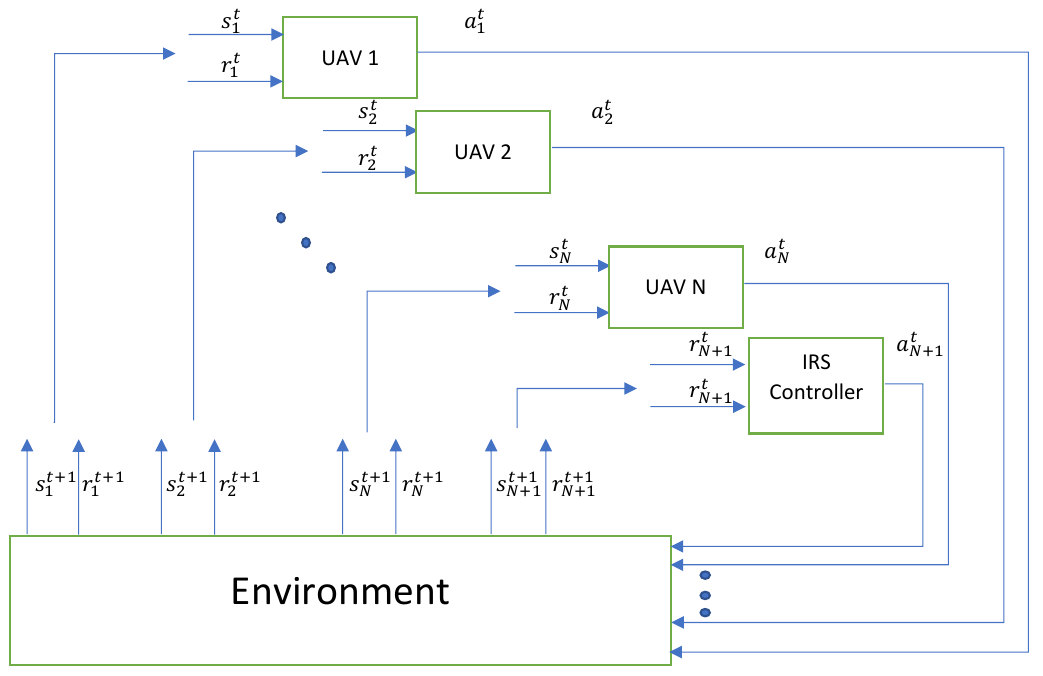}}
	\caption{A multi-agent learning for the RIS-assisted wireless networks}
	\label{fig:Parallel}
\end{figure}

In time step $t$, the $n$th UAV decides the transmit power $P_n$ and the RIS chooses the proper phase-shift matrix $\Phi^t$ at the state $s^t$ for maximising the EE performance. In particular, our parallel model is described as in Fig. \ref{fig:Parallel}. The UAV and the RIS have the local information and interact with the environment to search for an optimal policy $\pi^*$. The agents at each timestep choose and execute the action toward the environment. Then, the environment will respond by a value of reward toward the agents. Based on the responsed reward, the agents adjust the value of parameters in the action-chosen scheme for finding an optimal policy $\pi^*$. The details of our proposed techniques for joint optimisation of power allocation at the UAV and phase-shift matrix at the RIS are described in Algorithm~\ref{alg:PDDPG}. The agent $N+1$ represents the RIS controller.

 \begin{algorithm}[t!]
	\caption{Parallel learning for joint power allocation and phase-shift matrix in RIS-assisted UAV communications.}
	\begin{algorithmic}[1]
		\label{alg:PDDPG}
		\FOR{Agent $\varpi = 1, \dots, N, N+1$}
		\STATE Initialise the critic network $Q_\varpi(s, a; \theta_q)$, the target critic networks $Q'_\varpi$ and actor network $\mu_\varpi(s; \theta_\mu)$, target actor network $\mu'_\varpi$  for the agent $\varpi$
		\STATE Initialise replay memory pool $\mathcal{D}_\varpi$  for the agent $\varpi$
		\ENDFOR
		\FOR{episode = $1,\dots, E$}
		\STATE Initialise an action exploration process $\mathcal{N}$
		\STATE Receive initial observation state $s^0$
		\FOR{iteration = $1,\dots, T$}
		\FOR{Agent $\varpi = 1, \dots, N, N+1$}
		\STATE Execute the action $a_\varpi^t$ obtained at state $s^t$
		\STATE Update the reward $r^t_\varpi$ according to (\ref{equ:reward})
		\STATE Observe the new state $s_\varpi^{t+1}$
		\STATE Store transition $(s_\varpi^t, a_\varpi^t, r_\varpi^t, s_\varpi^{t+1})$ into replay buffer $\mathcal{B}_\varpi$
		\STATE Sample randomly a mini-batch of $D$ transitions $(s_\varpi^i, a_\varpi^i, r_\varpi^i, s_\varpi^{i+1})$ from $\mathcal{B}_\varpi$
		\STATE Update critic parameter by SGD using the loss Equ. (\ref{equ:loss})
		\STATE Update the actor policy parameter Equ. (\ref{equ:updateActor})
		\STATE Update the target networks as in (\ref{equ:updatePar1}) and (\ref{equ:updatePar2})
		\STATE Update the state $s^t_\varpi = s^{t+1}_\varpi$
		\ENDFOR
		\ENDFOR
		\ENDFOR
	\end{algorithmic}
\end{algorithm}

\section{Proximal Policy Optimisation for Centralised and Decentralised Problem.}\label{Sec:PPO}
Instead of using a hybrid model for continuous action space as in the DDPG algorithm, we propose an on-policy algorithm, namely proximal policy optimisation (PPO), with an efficient learning technique to achieve a better performance. In the PPO algorithm, we compare the current policy and obtained policy to find maximisation of the objective function as
\begin{equation}
\begin{split}
\mathcal{L} (s, a; \theta) &= \mathbb{E} \Bigg[ \frac{\pi(s, a; \theta)}{\pi (s, a; \theta_{old})} A^\pi(s, a) \Bigg]\\
&=\mathbb{E} \Bigg[  p^t_\theta A^\pi(s, a) \Bigg],
\end{split}
\end{equation}
where $p^t_\theta = \frac{\pi(s, a; \theta)}{\pi (s, a; \theta_{old})}$ denote the probability ratio and $A^\pi (s, a) = Q^\pi(s,a) - V^\pi(s)$ is an estimator of the advantage function defined in \cite{JS:16:ICLR}. We use SGD for training networks with a mini-batch $D$ to maximise the objective. Thus, the policy is updated by
\begin{equation}
\label{equ:policyPPO}
\theta^{t+1} = \argmax \mathbb{E} \Big[\mathcal{L}(s, a ; \theta^t)\Big].
\end{equation}
In this work, we use the clipping method function $\clip(p^t_\theta, 1-\epsilon, 1+\epsilon )$ for limiting the objective value to avoid the excessive modification as follows  \cite{JS:16:ICLR}:
\begin{equation}
\begin{split}
\mathcal{L}^{\mathsf{CLIP}} (s, a; \theta) = \mathbb{E} \Bigg[ & \min \Big(p^t_\theta A^\pi(s, a), \\& \clip(p^t_\theta, 1-\epsilon, 1+\epsilon )A^\pi (s, a)\Big) \Bigg],
\end{split}
\end{equation}
where $\epsilon$ is a small constant. We use the upper bound with $1+\epsilon$ when the advantage $A^\pi(s,a)$ is positive. In this case, the objective is equal to
\begin{equation}
\mathcal{L}^{\mathsf{CLIP}} (s, a; \theta) = \min \Bigg( \frac{\pi(s, a; \theta)}{\pi (s, a; \theta_{old})}, (1+\epsilon)  \Bigg) A^\pi(s,a).
\end{equation}

While the advantage $A^\pi (s,a)$ is positive, the minimum term puts a ceiling to the increased objective. Once $\pi (s, a; \theta) > (1+\epsilon) \pi (s, a; \theta_{old})$, the objective is limited by $(1+ \epsilon) A^\pi (s,a)$. Similarly, when the advantage is negative, the objective can be written as follows:
\begin{equation}
\mathcal{L}^{\mathsf{CLIP}} (s, a; \theta) = \max \Bigg( \frac{\pi(s, a; \theta)}{\pi (s, a; \theta_{old})}, (1-\epsilon)  \Bigg) A^\pi(s,a).
\end{equation}

When the advantage is negative, if $\pi(s,a; \theta)$ decreases the objective will increase. Thus, the maximum term puts a ceiling and once $\pi (s, a; \theta) < (1-\epsilon) \pi(s, a; \theta_{old})$, the objective is limited by  $(1- \epsilon) A^\pi (s,a)$. These clipping surrogate methods restrict the new policy not going far from the old policy.

Furthermore, we use an advantage function $A^\pi(s, a)$ as follows \cite{Mnih:16}:
\begin{equation}
\label{equ:A}
A^\pi(s, a) = r^t + \zeta V^\pi(s^{t+1}) -V^\pi(s^t)
\end{equation}

\section{Simulation Results}\label{Sec:Results}

For implementing our algorithms, we use the Tensorflow 1.13.1 \cite{Abadi:16}. We deploy $N = 3$ UAVs to serve $3$ clusters at the fixed location $(0, 0,200), (200, 300, 200), (400, 0, 200)$. We assume $d/\lambda = 1/2$. The total power consumption at the RIS and non-transmit power of UAV is set to $P_K + P_c = 4$W. For the neural network setting, in the DDPG algorithm, we use learning rate $lr1 = 0.001$ and $lr2 =0.002$ for the actor and critic network, respectively. In the PPO algorithm, we use the learning rate $lr= 0.00001$. Other parameters are provided in Table \ref{tab:Params}. In this section, the four proposed schemes in previous sections are summarised as follows:

\begin{itemize}
	\item \textbf{Our centralised DDPG algorithm (C-DDPG)}: As we explained in Section \ref{Sec:DDPG}, we use the DDPG algorithm for jointly optimising the transmit power of the UAV and the phase-shift matrix of the RIS in a centralised manner.
\item \textbf{Parallel learning for the DDPG method (P-DDPG)}: We consider parallel learning to help to reduce the information transmission delay and errors while ensuring the network performance.
\item \textbf{Our centralised PPO algorithm (C-PPO)}: Instead of using the DDPG algorithm, we use the PPO algorithm for solving the centralised problem.
\item \textbf{Parallel learning for the PPO algorithm (P-PPO)}: We also deploy the PPO algorithm for parallel learning in our joint power allocation and phase-shift matrix optimisation in multi-UAV and RIS-assisted wireless networks.
\end{itemize}
In addition, to highlight the advantage of our proposals, we also compare our four proposed methods with the following schemes:
\begin{itemize}
	\item \textbf{Max power transmission (MPT)}: We use the maximal transmit power at the UAV and optimise the phase-shift of the RIS by using the PPO algorithm.
\item \textbf{Random selection scheme (RSS)}: We select randomly the phase-shift at the RIS and optimise the transmit power at the UAV.
\end{itemize}

\begin{table}[h!]
	\renewcommand{\arraystretch}{1.2}
	\caption{SIMULATION PARAMETERS}
	\label{tab:Params}
	\centering
	\begin{tabular}{l|l}
		\hline
		Parameters & Value \\
		\hline
		Bandwidth ($W$)  & $1$ MHz \\
		UAV transmission power & $5$ W \\
		UAV's coverage & $500$ m\\
		The RIS's position & $(500, 500, 30)$\\
		Path-loss parameter & $\kappa_1 = 2, \kappa_2 = 2.2$\\
		Channel power gain & $\beta_0 = -30$ dB\\
		Rician factor & $\beta_1 = 4$\\
		Noise power & $\alpha^2 = -134$ dBm\\
		Discounting factor & $\zeta = 0.9$\\
		Max number of UEs & $30$\\
		Initial batch size & $ D = 32$ \\
		
		\hline
	\end{tabular}
\end{table}

\begin{figure}[h!]
	\centering
	\subfigure{\includegraphics[width=0.5\textwidth]{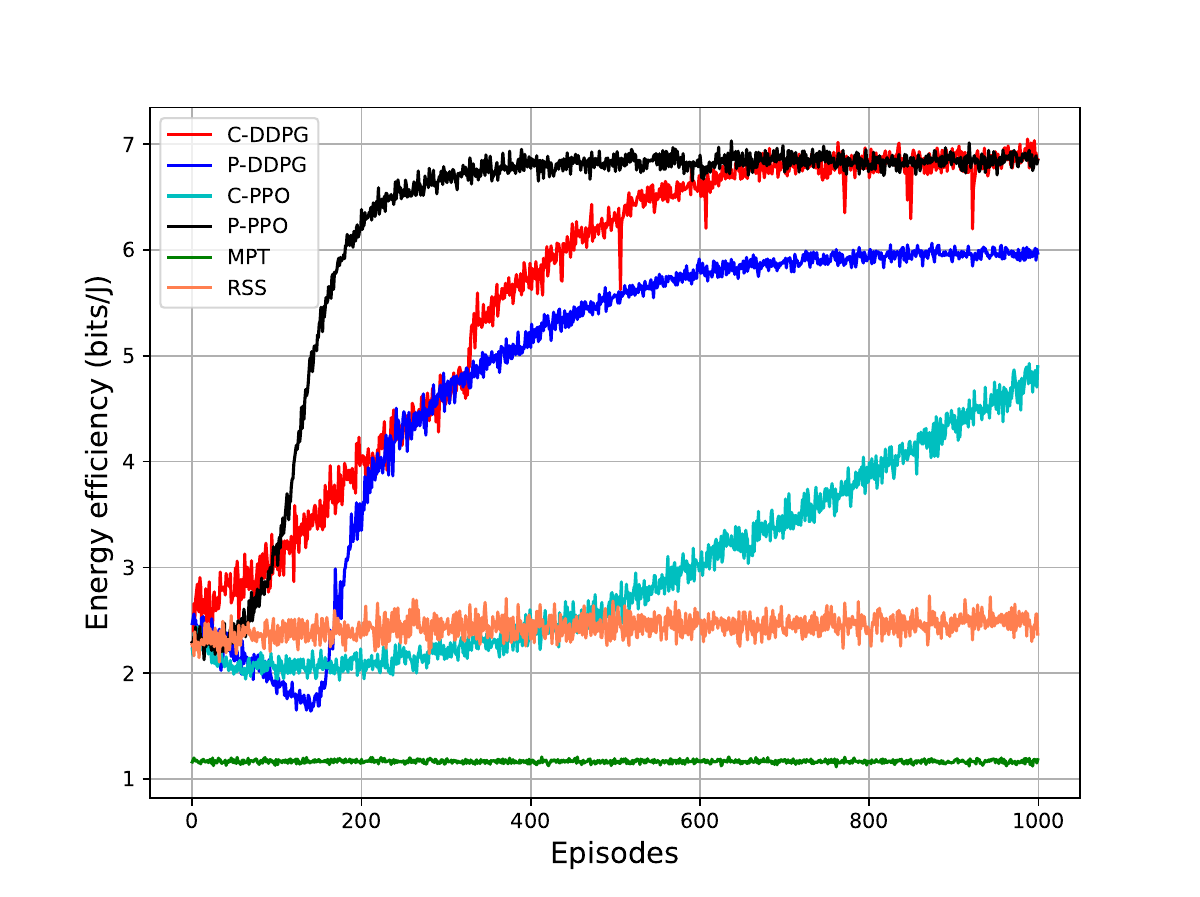}}
	\caption{The EE with $M = 10$ and $K = 20$.}
	\label{fig:Reward}
\end{figure}

In Fig.~\ref{fig:Reward}, we show the EE performance of our proposed method in both centralised and decentralised learning with $M = 10$ and $K = 20$. The methods based on parallel learning reach the best results with the P-DDPG and P-PPO algorithm. Both are higher than the ones using the C-DDPG and C-PPO algorithm in the centralised learning. The convergence of the P-PPO is fastest and following by the P-DDPG algorithm. As can be observed from this figure our proposed scheme with joint optimisation using the DRL techniques outperform the other approaches using the MPT and RSS methods.

\begin{figure}[h!]
	\centering
	\subfigure{\includegraphics[width=0.5\textwidth]{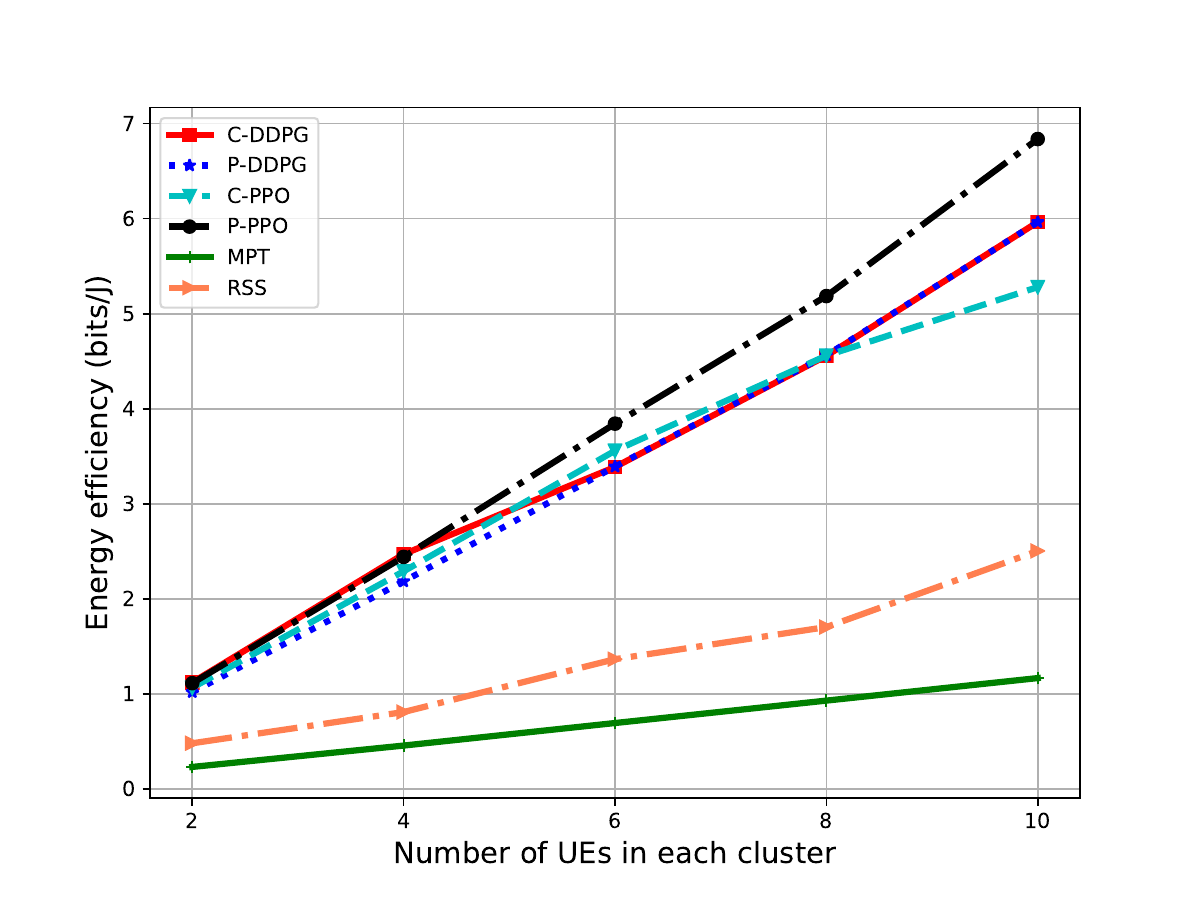}}
	\caption{The EE versus the number of UEs in each cluster, $M$.}
	\label{fig:NoUEs}
\end{figure}

In Fig.~\ref{fig:NoUEs}, the EE performance of our methods in comparison with other baseline schemes are presented with the different number of UEs in each cluster, $M$, for the number of RIS elements $K=20$. Again, the P-PPO method shows better EE performance than the centralised C-PPO and the ones using the C-DDPG algorithm. The MPT and RSS method are less effective for the joint power allocation and phase shift matrix optimisation in the UAV-assisted wireless network with the support of the RIS.

\begin{figure}[h!]
	\centering
	\subfigure{\includegraphics[width=0.5\textwidth]{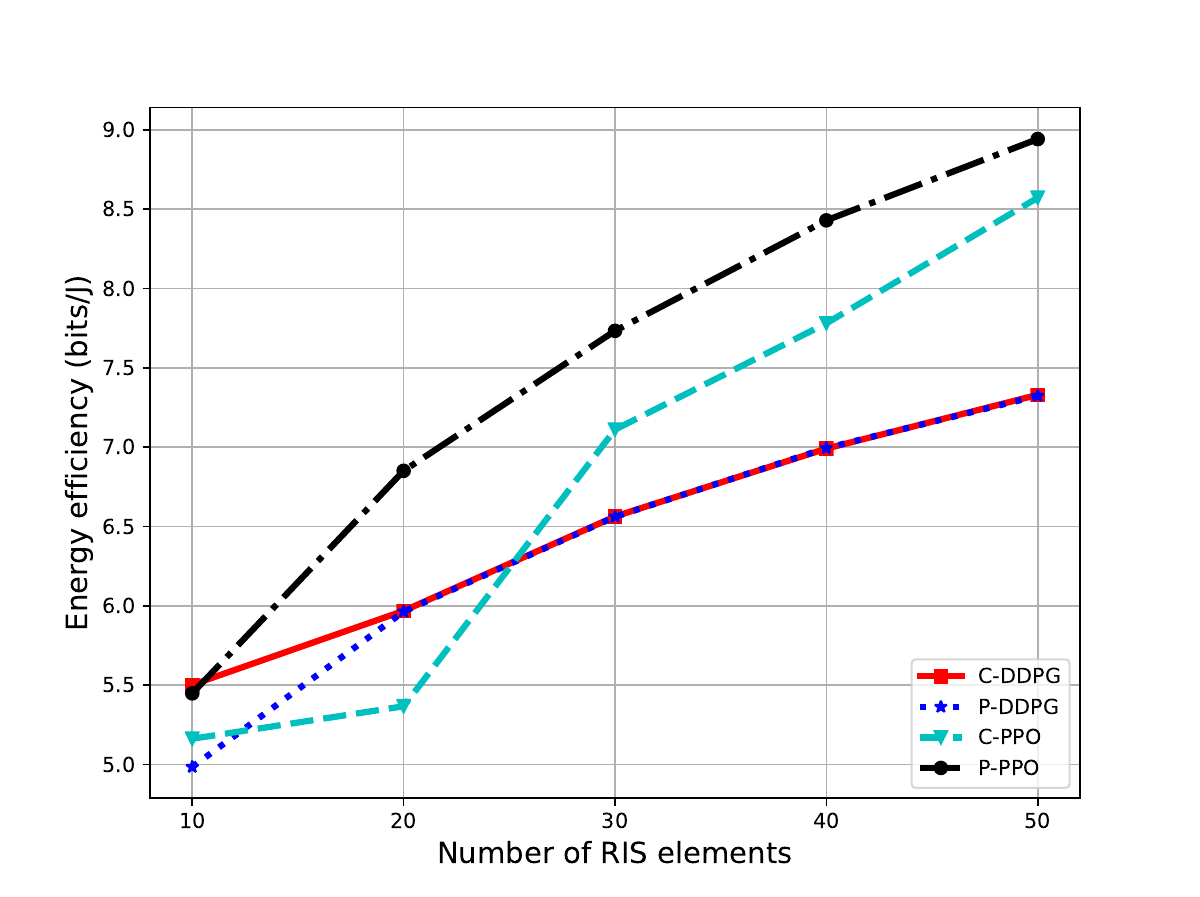}}
	\caption{The EE versus the number of the RIS elements, $K$.}
	\label{fig:NoIRS}
\end{figure}

In Fig.~\ref{fig:NoIRS}, we plot the EE performance versus the number of the RIS elements ($K$) when the number of UEs in each cluster equals to ten ($M=10$). We achieve the best EE performance with the P-PPO algorithm despite the value of $K$. When the number of RIS elements becomes higher (e.g., $K>25$), the methods based on the C-PPO algorithm are more effective than the ones using the DDPG algorithm. In contrast, for a smaller value of $K$,  the methods based on the C-DDPG algorithm are better than the centralised learning with the C-PPO algorithm. For all values of $K$, the best performance can be achieved with P-PPO algorithm, which demonstrates the fact that the P-PPO algorithm is stable and practical for every environmental setting under the joint optimisation of power allocation at UAVs and the phase-shift matrix at RIS.

\begin{figure}[h!]
	\centering
	\subfigure{\includegraphics[width=0.5\textwidth]{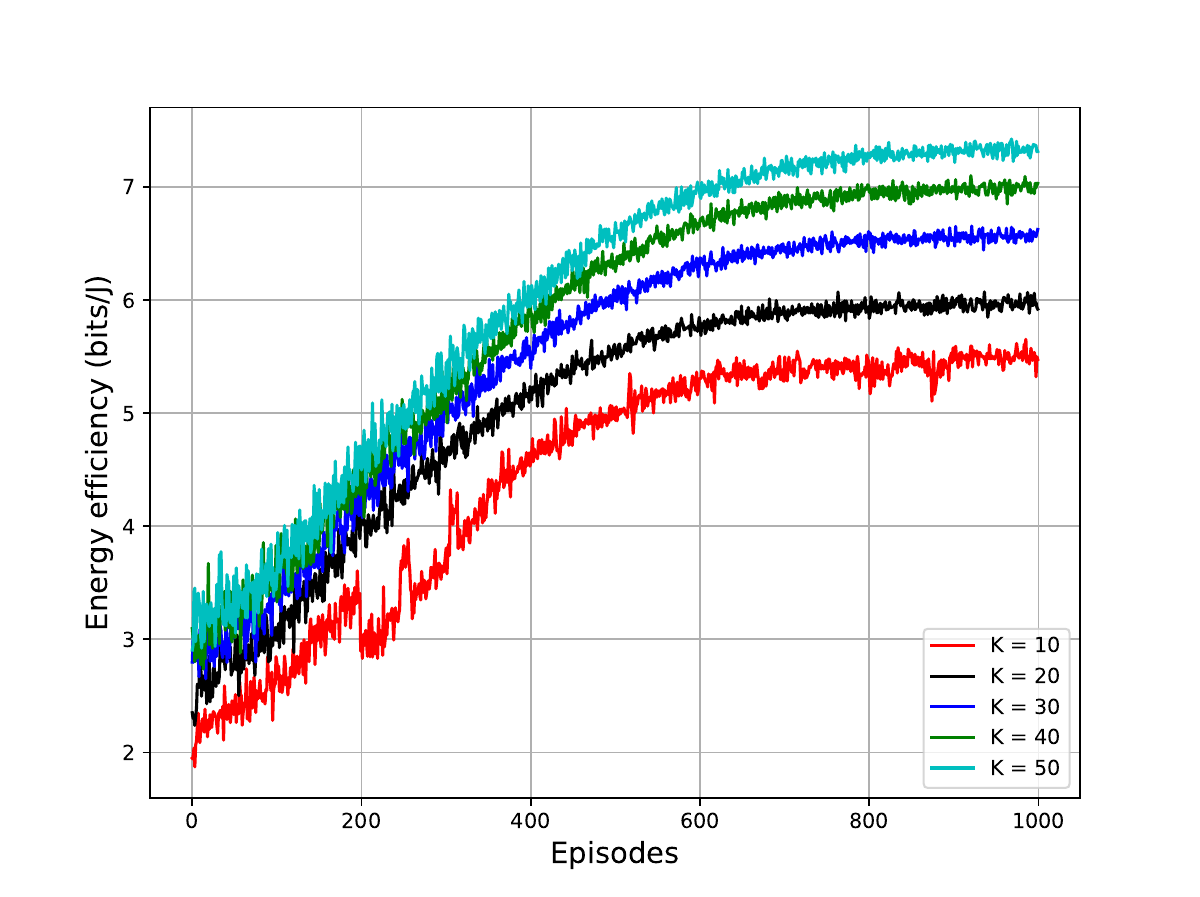}}
	\caption{The EE of the C-DDPG algorithm with different number of the RIS elements, $K$.}
	\label{fig:NoIRS_DDPG}
	
	\end{figure}

\begin{figure}[h!]
	\centering
	\subfigure{\includegraphics[width=0.5\textwidth]{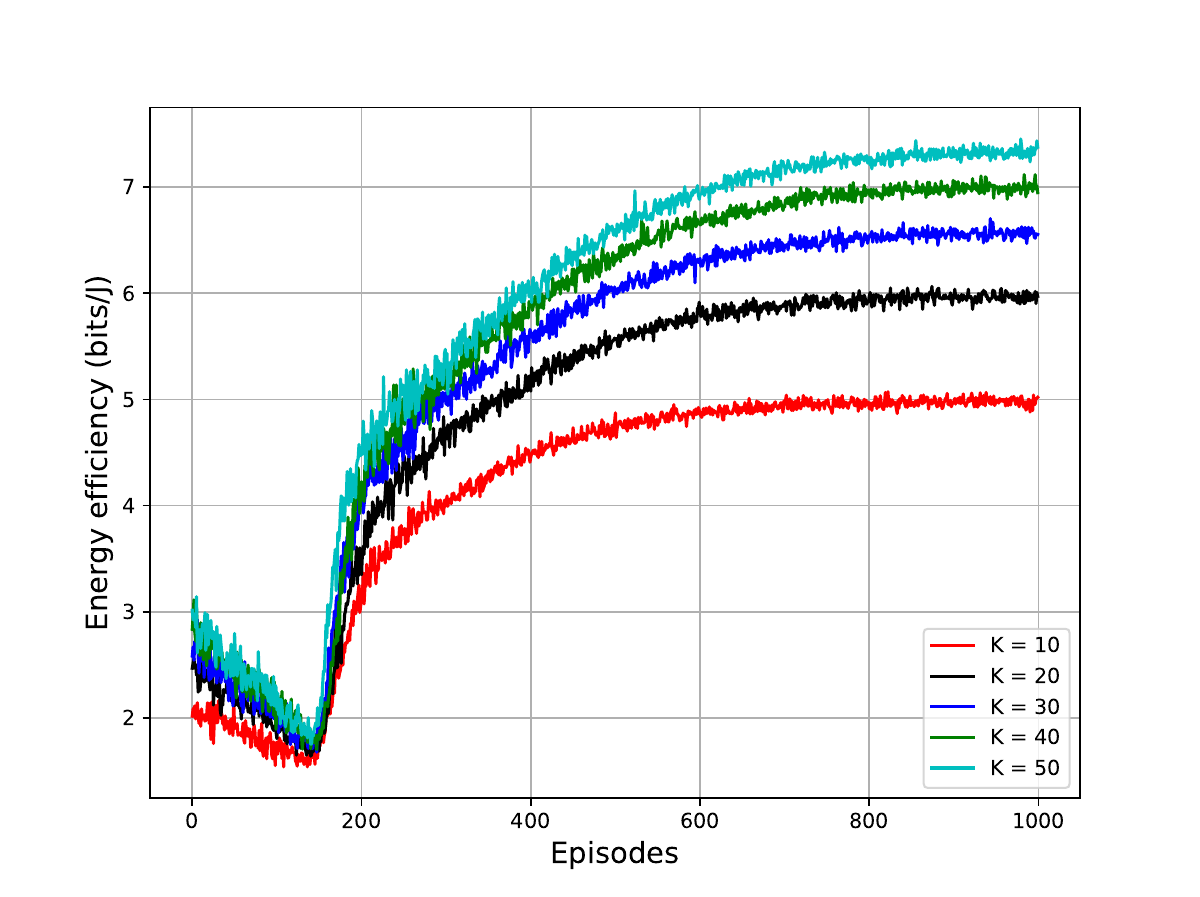}}
	\caption{The EE of the P-DDPG algorithm with different number of the RIS elements, $K$.}
	\label{fig:NoIRS_PDDPG}
\end{figure}

The EE performances of the DDPG algorithm versus episodes for different number of RIS elements using the centralised learning and parallel learning are shown in Fig.~\ref{fig:NoIRS_DDPG} and Fig.~\ref{fig:NoIRS_PDDPG}, respectively. With the higher number of RIS elements, the performance increase while the convergence rate is still similar for both centralised and parallel approaches. The result converges after about $600$ episodes when the \emph{exploration} is set to $3$ and $\psi = 0.99995$. Thus, depending on the specific purpose, we can deploy the configurable RIS with fast learning.


\begin{figure}[h!]
	\centering
	\subfigure{\includegraphics[width=0.5\textwidth]{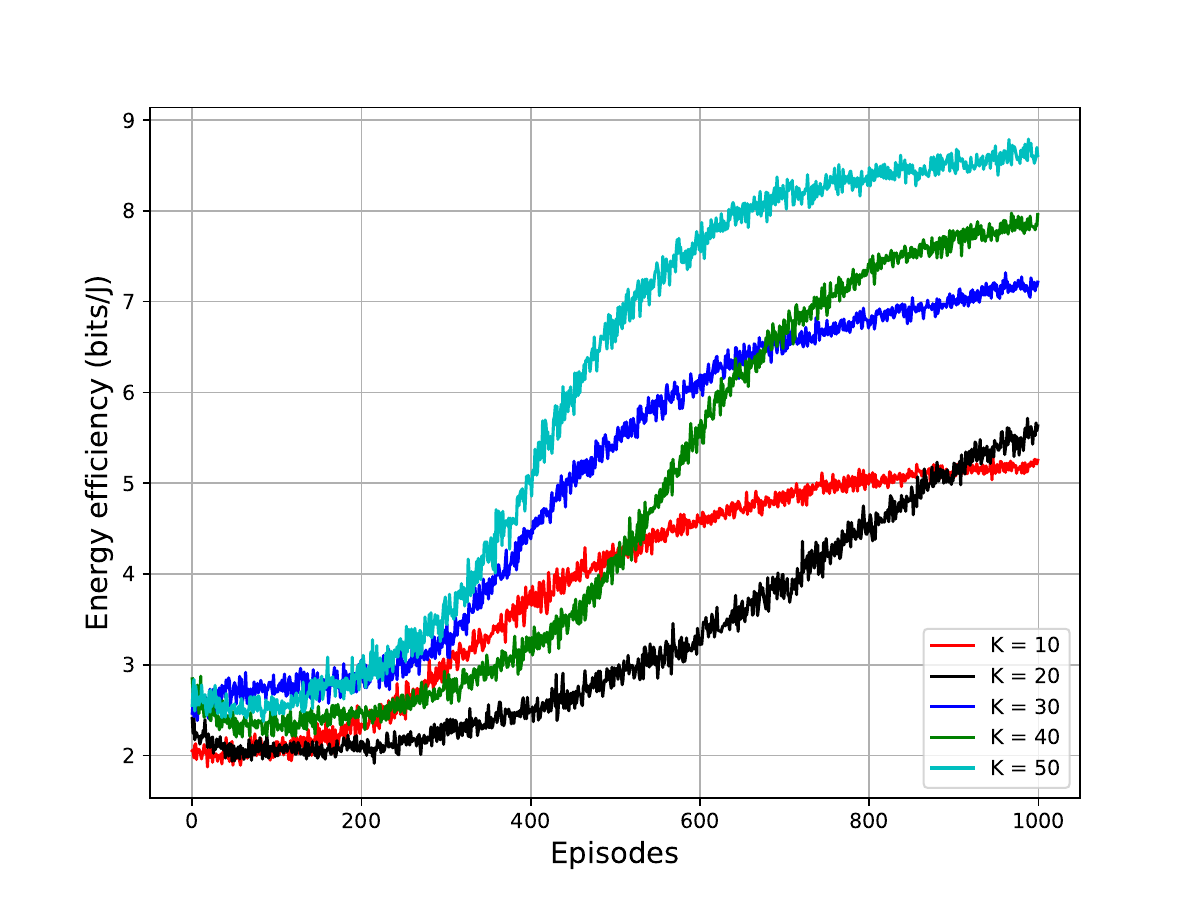}}
	\caption{The EE of the C-PPO algorithm with different number of the RIS elements, $K$.}
	\label{fig:NoIRS_PPO}
\end{figure}

\begin{figure}[h!]
	\centering
	\subfigure{\includegraphics[width=0.5\textwidth]{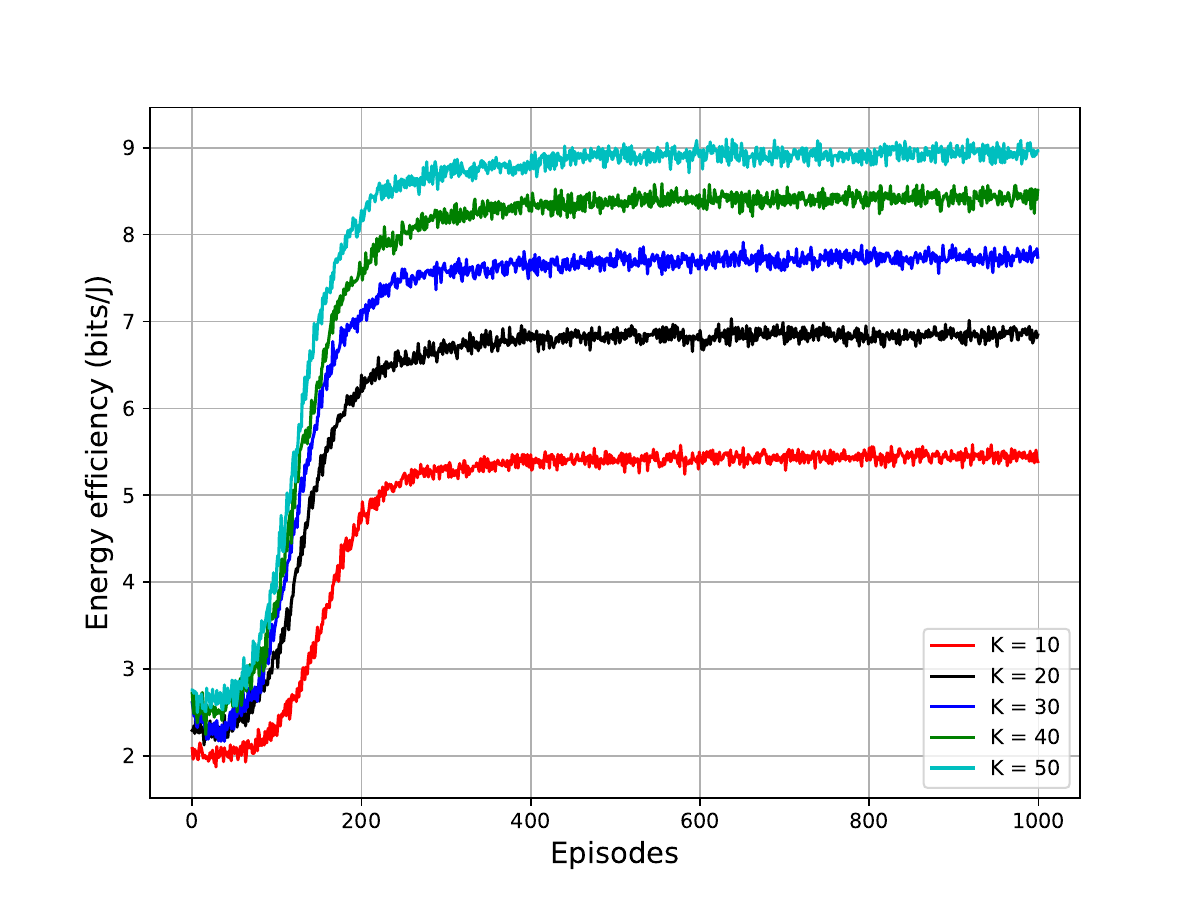}}
	\caption{The EE of the P-PPO algorithm with different number of the RIS elements, $K$.}
	\label{fig:NoIRS_PPPO}
\end{figure}

Similarly, the EE performance of PPO algorithm versus episodes for different number of RIS elements using the centralised learning and parallel earning are plotted in Fig.~\ref{fig:NoIRS_PPO} and Fig.~\ref{fig:NoIRS_PPPO}, respectively. While the performance using centralised approach (C-PPO) is unstable and takes around $800$ episodes for convergence, the parallel approach (P-PPO algorithm) shows a solid performance even when increasing the number of the RIS elements. The convergence for P-PPO is still stable and even faster with the higher number of RIS elements. We need only about $200$ episodes for convergence. Furthermore, we use neural networks for the DDPG and PPO algorithm; thus, the system can be easily deployed after training and the agent can choose the action immediately.

\section{Conclusions}\label{Sec:Con}
In this paper, we have proposed multi-UAV networks supported by a RIS panel to enhance the network performance. To maximise the EE of the considered networks, the transmit power at the UAV and the phase-shift matrix at the RIS were jointly optimised by using the DDPG method and PPO technique in a centralised approach. Moreover, to reduce the network's delay and the power for exchanging the information, we proposed parallel learning for the optimisation problem. The results suggested that we can deploy the DRL algorithms for the real-time optimisation with impressive results compared to other baseline schemes. For the future work, we will improve the model with multiple RIS panel and cooperative communications with an fully autonomous ability in the futures.

\section*{Acknowledgement}
This work was supported in part by the U.K. Royal Academy of Engineering (RAEng) under the RAEng Research Chair and Senior Research Fellowship scheme Grant RCSRF2021$\backslash$11$\backslash$41.

\bibliographystyle{IEEEtran}

\bibliography{IEEEabrv,reference}

\end{document}